\documentclass[sn-mathphys,Numbered]{sn-jnl}

\usepackage{makecell}
\usepackage{graphicx}%
\usepackage{multirow}%
\usepackage{amsmath,amssymb,amsfonts}%
\usepackage{amsthm}%
\usepackage{mathrsfs}%
\usepackage[title]{appendix}%
\usepackage{xcolor}%
\usepackage{textcomp}%
\usepackage{manyfoot}%
\usepackage{booktabs}%
\usepackage{algorithm}%
\usepackage{algorithmicx}%
\usepackage{algpseudocode}%
\usepackage{listings}%
\usepackage{pgf}
\usepackage{lmodern}
\usepackage[shortlabels]{enumitem}
\usepackage{import}
\usepackage{todonotes}
\usepackage{tikz}
\usetikzlibrary{quantikz2}
\usetikzlibrary{backgrounds}
\usetikzlibrary{shapes.geometric, arrows}
\tikzstyle{arrow} = [thick, ->, >=latex]

\usepackage{anyfontsize}

\newcommand{\mathdefault}[1][]{}

\usepackage{etoolbox}   
\makeatletter
\patchcmd{\@maketitle}{\artauthors}{{\artauthors}}{}{}
\makeatother

\begin{document}

\title[Scalable General Error Mitigation for Quantum Circuits]{Scalable General Error Mitigation for Quantum Circuits}


\author[1]{\fnm{Philip} \sur{Döbler}}
\author[1]{\fnm{Jannik} \sur{Pflieger}}
\author[2]{\fnm{Fengping} \sur{Jin}}
\author[2]{\fnm{Hans} \sur{De Raedt}}
\author[2,3]{\fnm{Kristel} \sur{Michielsen}}
\author[2,1]{\fnm{Thomas} \sur{Lippert}}
\author*[1]{\fnm{Manpreet Singh} \sur{Jattana}}\email{jattana@em.uni-frankfurt.de}

\affil[1]{\orgdiv{Modular Supercomputing and Quantum Computing}, \orgname{Goethe University Frankfurt}, \orgaddress{\street{Kettenhofweg 139}, \city{Frankfurt am Main}, \postcode{60325}, \country{Germany}}}
\affil[2]{\orgdiv{Jülich Supercomputing Centre, Institute for Advanced Simulation}, \orgname{Forschungszentrum Jülich}, \orgaddress{\street{Wilhelm-Johnen-Straße}, \city{Jülich}, \postcode{52428}, \country{Germany}}}
\affil[3]{\orgname{RWTH Aachen University}, \orgaddress{\street{Templergraben 55}, \city{Aachen}, \postcode{52062}, \country{Germany}}}


\abstract{In quantum computing, error mitigation is a method to improve the results of an error-prone quantum processor by post-processing them on a classical computer. In this work, we improve the General Error Mitigation (GEM) method for scalability. GEM relies on the use of a matrix to represent the device error, which requires the execution of $2^{n+1}$ calibration circuits on the quantum hardware, where $n$ is the number of qubits. With our improved method, the number of calibration runs is independent of the number of qubits and depends only on the number of non-zero states in the output distribution. We run 1853 randomly generated circuits with widths between 2-7 qubits and depths between 10-140 gates on IBMQ superconducting devices. The experiments show that the mitigation works comparably well to GEM, while requiring a fraction of the calibration runs. Finally, an experiment to mitigate errors in a 100 qubit circuit demonstrates the  scalable features of our method.}




\maketitle
\section{Introduction}
\label{sec_introduction}

Since the famous keynote of Richard Feynman in 1982 about the simulation of quantum physics with computers \cite{feynman_keynote}, there has been an increasing interest and effort in the field of quantum computing. For various problems, it has been shown that there is a huge potential for speed up over classical algorithms \cite{deutsch_josza, shors_algo, grover, quantum_sim, hhl}. However, state-of-the-art quantum computers are still error-prone, which prevents them from showing quantum advantage for real-world problems.

These error-prone devices are also called \emph{noisy intermediate-scale quantum (NISQ)} devices \cite{nisq_era}. To tackle noise, several \emph{error correction} codes have been proposed \cite{error_cor_shor, error_cor_steane, error_correction_laflamme}, but they all require multiple physical qubits to create one logical error-corrected qubit. The qubits on most NISQ devices are already a scarce resource, so reducing the number further would decrease the potential problem size. Due to the difficulty of manufacturing error-corrected devices, a new field of research called \emph{error mitigation} has emerged focusing on improving the quality of results from error-prone devices. There are two recent overview papers on the topic~\cite{error_mit_overview_1, error_mit_overview_2}. Both papers cluster the error mitigation methods into different categories, which we briefly mention below.

One category is \emph{zero-noise extrapolation}, also known as \emph{error extrapolation}. These methods artificially increase the noise for a certain circuit to subsequently extrapolate the result to zero noise. The concept was originally introduced in References~\cite{error_mit_exrapolation_1} and~\cite{error_mit_extrapolation_2} and refined in subsequent works~\cite{ error_mit_extrapolation_3, error_mit_extrapolation_4, error_mit_extrapolation_5, error_mit_extrapolation_6, error_mit_extrapolation_7, error_mit_extrapolation_8, error_mit_extrapolation_9, error_mit_extrapolation_10, error_mit_extrapolation_11, error_mit_extrapolation_12, error_mit_extrapolation_13, error_mit_extrapolation_14, error_mit_extrapolation_15}. With zero-noise extrapolation, it is important to choose the right extrapolation method and find a way to artificially increase the errors. For the latter, we can either add gates to the circuit or stretch the pulses that implement the gates. Adding gates has the advantage that we do not require low-level control of the hardware. However, we need detailed knowledge of the device's error model or bear the risk that the error model changes with the inserted gates and therefore changes the errors in an unintended way~\cite{error_mit_overview_1, error_mit_overview_2}.

Another frequently used category is \emph{probabilistic error cancellation}. Here we aim to express the error-free expectation value as a linear combination of expectation values that can be sampled from error-prone quantum circuits~\cite{error_mit_overview_2}. The method was introduced in the same paper as the zero-noise extrapolation~\cite{error_mit_extrapolation_2} and expanded and improved in many other works~\cite{error_mit_extrapolation_3, error_mit_extrapolation_9, error_mit_pec_1, error_mit_pec_2, error_mit_pec_3, error_mit_pec_4, error_mit_pec_5, error_mit_pec_6, error_mit_pec_7, error_mit_pec_8, error_mit_pec_9, error_mit_pec_10, error_mit_pec_11}. The downsides of this method are that we need detailed knowledge of the noise present in the device and a sampling overhead which grows exponentially with the error rate of the circuit~\cite{error_mit_overview_2}.

Since in error mitigation we are trying to map the error-prone output of quantum computers to an error-free output, it is natural to use machine learning methods that learn this mapping. Such methods have been proposed and tested~\cite{error_mit_ml_1, error_mit_ml_2, error_mit_ml_3, error_mit_ml_4, error_mit_ml_5, error_mit_ml_6, error_mit_ml_7, error_mit_ml_8, error_mit_ml_9}. For the training data, it is necessary to have circuits that are subject to similar errors as the original circuit, but can be simulated classically. For this purpose we can replace all or most gates in a circuit by Clifford gates~\cite{error_mit_ml_6}, which can be efficiently simulated~\cite{gottesman_knill}. We can consider the training either as part of the application or, if the training set is large enough to be generalized to different applications, as part of the device calibration~\cite{error_mit_overview_2}. In both cases, the training comes with a huge sampling overhead.

A fourth category are matrix-based methods. With these methods we create what is called a \emph{complete assignment matrix}~\cite{error_mit_spam_3}, which represents the errors in the device. From the measured frequencies, we can calculate the mitigated frequencies using this matrix. Reference~\cite{error_mit_overview_1} lists these methods under ``mitigating measurement error'', since they were originally used to mitigate state preparation and measurement (SPAM) errors~\cite{error_mit_spam_1, error_mit_spam_2}, but no gate errors. One problem with matrix-based methods is that $2^n$ calibration circuits are needed to construct the complete assignment matrix, where $n$ is the number of qubits. Reference~\cite{error_mit_spam_3} shows that scalability can be achieved by using iterative methods like Jacobi iteration~\cite{jacobi_iteration}, generalized-minimal-residual~\cite{gmres} or biconjugate-gradient-stabilized~\cite{bi-cgstab} method. The aforementioned mitigation methods do not take gate errors into account. However, References~\cite{error_mit_jattana} and~\cite{jattana_diss} extend the method to include gate errors as well, which they call \emph{general error mitigation (GEM)}. The method is relatively easy to use, as evidenced by the fact that it has been used in other works to improve results~\cite{citing_gem_1, citing_gem_2, citing_gem_3, citing_gem_4, citing_gem_5, citing_gem_6}.

In Reference~\cite{error_mit_jattana}, the authors list seven criteria, that an error mitigation method should fulfill: Result recovery to a satisfactory accuracy (1), independence from the depth of a circuit (2), not relying on the knowledge of the device's error model (3), practical implementability (4), no need for additional quantum hardware (5), independence from the gate set (6) and no knowledge about the output of the circuit (7). The GEM method fulfills the Criteria~2, 3 and 5-7. For Criterion~1 users must decide if the result recovery is sufficient for their application. We cannot say that the method is practically implementable (Criterion~4) because of the exponential number of calibrations required. It can only be used for small devices. However, there are already machines from IBMQ with 127 qubits available over the cloud~\cite{ibm_devices} and we can expect to see even larger devices in the future.

To ensure that matrix-based mitigation methods can be applied to larger devices, we present in this paper an extension to the GEM method that solves the scalability problem formulated in Criterion~4. We refer to this method as \emph{scalable general error mitigation (SGEM)}. The main contributions of this paper are:
\begin{itemize}
    \item A scalable heuristic error mitigation scheme based on matrices suitable for large qubit numbers.
    \item Evaluation of the method with randomized circuits on superconducting devices.
    \item Large scale testing with up to 100 qubit circuits.
\end{itemize}

The paper is structured as follows: In Section~\ref{sec_matrix_based} we introduce a matrix-based error mitigation method in more detail, in Section~\ref{sec_sparse_matrix} we show how this method can be made scalable to large problem sizes. Section~\ref{sec_experimental_results} contains our experimental results. We conclude the paper in Section~\ref{sec_conclusion}.
\section{Matrix-based error mitigation}
\label{sec_matrix_based}

For matrix-based error mitigation we construct a matrix $M_I$ such that
\begin{equation}
    M_IE = V,
\end{equation}
where $V$ is a vector of length $2^n$ containing the frequencies $v_i$ produced by an error-prone device and $E$ is a vector of length $2^n$ with the frequencies $e_i$ an ideal device would give with an infinite number of shots. Others have called $M_I$ \emph{complete assignment matrix}~\cite{error_mit_spam_3}. It maps the ideal frequencies to the frequencies of the error-prone device. If the device has no errors, $M_I$ is the identity matrix of size $2^n \times 2^n$. For increasing amount of errors, the diagonal values get smaller and the off-diagonal values increase.

Assuming we can construct the mitigation matrix $M_I$, we can invert $M_I$ to calculate the exact frequencies $e_i$ from the measured frequencies $v_i$:

\begin{equation}
    E = M_I^{-1}V.
\end{equation}

There are however two problems: There is no known method to construct $M_I$ directly, there are only methods to get an approximation of the ideal $M_I$, which we call \emph{approximated complete assignment matrix} $M_A$. Therefore, we do not obtain the exact frequencies $E$, but something we call \emph{mitigated frequencies} denoted by the vector $X$ with elements $x_i$. We postulate that $X \to E$, as $M_A \to M_I$. Another problem is that $X$ must contain relative frequencies in the interval $[0,1]$ \cite{error_mit_spam_2}. If we invert $M_A$ to obtain $X$, there is no guarantee that the result is bound to the interval. To resolve this issue, we find the minimum of a cost function $f(x)$ with
\begin{equation}\label{equ_least_squares}
    f(x) = \sum_{i=1}^{2^n}(v_i - (M_A X)_i)^2
\end{equation}
and the constraints $0\leq x_i \leq 1$ and $\sum x_i = 1$ for all $i$.

\subsection{SPAM protocol}
\begin{figure}[htb]
    \centering
    \begin{tikzpicture}[node distance=0cm]
    \def\vspace{2};
    
    \node (circ_00) at (0, 0*\vspace) [anchor=north]{
    \begin{quantikz}
        \lstick{$\ket{0}$} & \meter{}\\
        \lstick{$\ket{0}$} & \meter{}
    \end{quantikz}};
    \node[right=of circ_00] (circ_00_label) {$\begin{pmatrix}0.90\\0.04\\0.04\\0.02\end{pmatrix}$};

    \node (circ_01) at (0, -1*\vspace) [anchor=north]{
    \begin{quantikz}
        \lstick{$\ket{0}$} & \qw & \meter{}\\
        \lstick{$\ket{0}$} & \gate{X} & \meter{}
    \end{quantikz}};
    \node[right=of circ_01] (circ_01_label) {$\begin{pmatrix}0.06\\0.85\\0.04\\0.05\end{pmatrix}$};

    \node (circ_10) at (0, -2*\vspace) [anchor=north]{
    \begin{quantikz}
        \lstick{$\ket{0}$} & \gate{X} & \meter{}\\
        \lstick{$\ket{0}$} & \qw & \meter{}
    \end{quantikz}};
    \node[right=of circ_10] (circ_10_label) {$\begin{pmatrix}0.03\\0.03\\0.92\\0.02\end{pmatrix}$};

    \node (circ_11) at (0, -3*\vspace) [anchor=north]{
    \begin{quantikz}
        \lstick{$\ket{0}$} & \gate{X} & \meter{}\\
        \lstick{$\ket{0}$} & \gate{X} & \meter{}
    \end{quantikz}};
    \node[right=of circ_11] (circ_11_label) {$\begin{pmatrix}0.04\\0.02\\0.05\\0.89\end{pmatrix}$};

    \node (matrix) at (6, -2*\vspace) {$M_A=\begin{pmatrix}0.90&0.06&0.03&0.04\\0.04&0.85&0.03&0.02\\0.04&0.04&0.92&0.05\\0.02&0.05&0.02&0.89\end{pmatrix}$};

    \draw [dashed] (-1.5,-\vspace-0.15) -- (3,-\vspace-0.15);
    \draw [dashed] (-1.5,-2*\vspace-0.15) -- (3,-2*\vspace-0.15);
    \draw [dashed] (-1.5,-3*\vspace-0.15) -- (3,-3*\vspace-0.15);
    \draw [arrow] (circ_00_label) to [out=0, in=90] ([xshift=-0.8cm]matrix.north);
    \draw [arrow] (circ_01_label) to [out=0, in=90] ([xshift=0cm]matrix.north);
    \draw [arrow] (circ_10_label) to [out=0, in=270] ([xshift=0.8cm]matrix.south);
    \draw [arrow] (circ_11_label) to [out=0, in=270] ([xshift=1.6cm]matrix.south);

\end{tikzpicture}
    \caption{Construction of the approximated complete assignment matrix for a device with two qubits. We prepare all possible basis states, for two qubits $\ket{00}$, $\ket{01}$, $\ket{10}$ and $\ket{11}$, and perform measurements. The resulting frequencies are the columns for the matrix. Since the device is noisy, the matrix is not the identity matrix but has off-diagonal entries.}
    \label{fig:circuits_simple}
\end{figure}
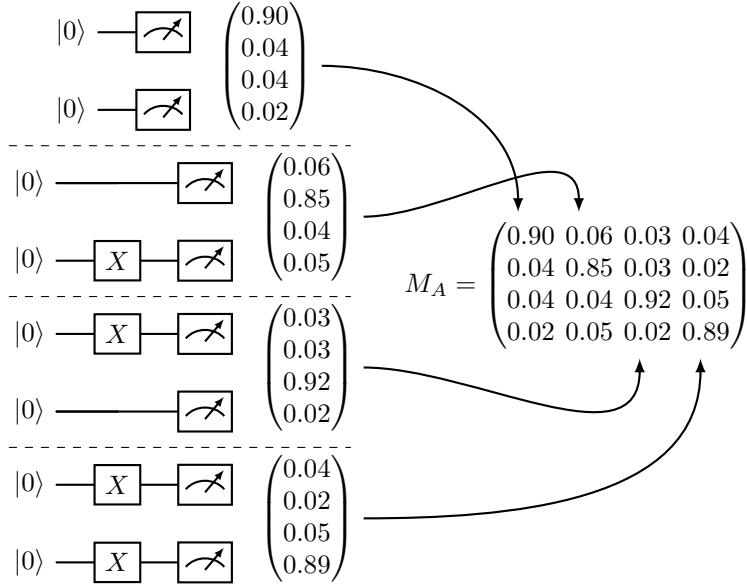

The above procedure leaves us with the challenge of constructing the matrix $M_A$ in a way that it comes as close as possible to $M_I$. In the simplest case, we prepare our system in all $2^n$ possible basis states and perform measurements. The frequencies of these calibrations are used as columns of the approximated complete assignment matrix. This procedure is shown in Figure~\ref{fig:circuits_simple}. 

As we only prepare the basis states and perform measurements, we can correct for state preparation and measurement (SPAM) errors, but not for gate errors. With non-trivial circuits, however the error is dominated by the gate errors and correcting only state preparation and measurement errors is insufficient. Another issue with this protocol is that we need to run $2^n$ additional circuits, which is not feasible for large qubit numbers, violating Criterion~4.

\subsection{General error mitigation}
\label{sec_general_mitigation}
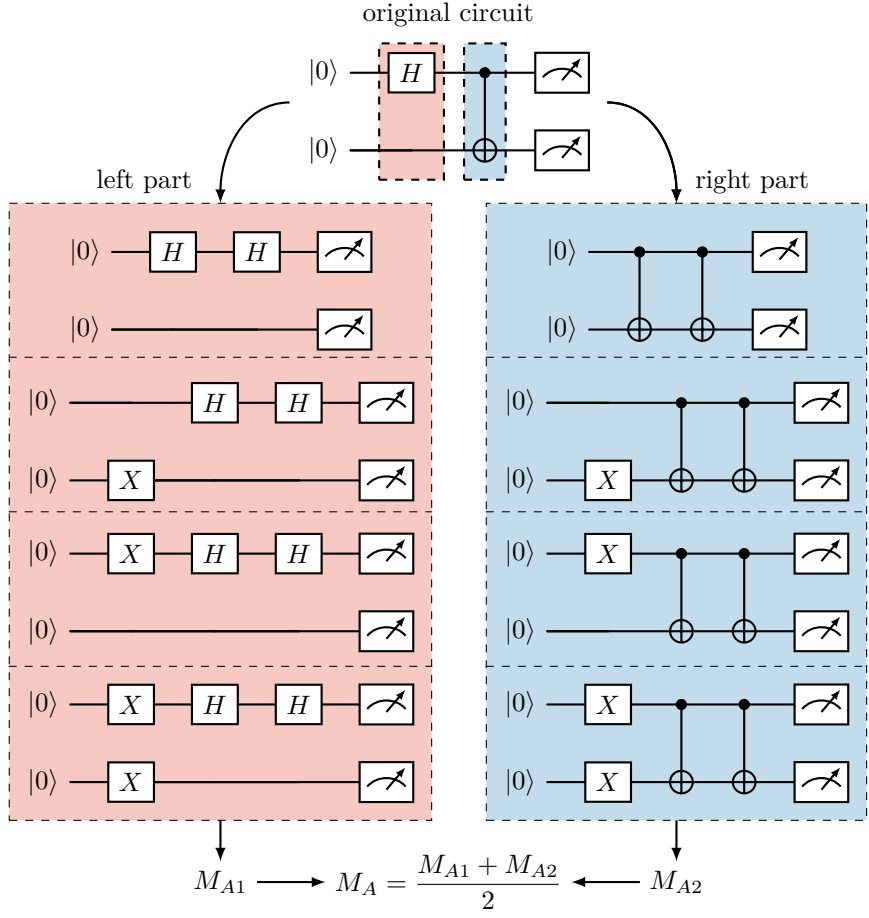
\begin{figure}[htb]
    \centering
    \definecolor{gnu_plot_1}{HTML}{E24A33}
\definecolor{gnu_plot_2}{HTML}{348ABD}

\begin{tikzpicture}
    \def\vspace{-2};
    \def\hspace{6};

    \node at (0.5*\hspace,0.5) {original circuit};
    \node (org) at (0.5*\hspace,0.5) [anchor=north]{
    \begin{quantikz}
        \lstick{$\ket{0}$} & \gate{H}\gategroup[2,steps=1,style={inner
sep=0pt,dashed,fill=gnu_plot_1!30},background]{} & \ctrl{1}\gategroup[2,steps=1,style={inner
sep=0pt,dashed,fill=gnu_plot_2!30},background]{} & \meter{}\\
        \lstick{$\ket{0}$} & \qw & \targ{} & \meter{}
    \end{quantikz}};
    
    \node (left_1) at (0,1*\vspace) [anchor=north]{
    \begin{quantikz}
        \lstick{$\ket{0}$} & \gate{H} & \gate{H} & \meter{}\\
        \lstick{$\ket{0}$} & \qw & \qw & \meter{}
    \end{quantikz}};

    \node at (0,2*\vspace) [anchor=north]{
    \begin{quantikz}
        \lstick{$\ket{0}$} & \qw & \gate{H} & \gate{H} & \meter{}\\
        \lstick{$\ket{0}$} & \gate{X} & \qw & \qw & \meter{}
    \end{quantikz}};

    \node at (0,3*\vspace) [anchor=north]{
    \begin{quantikz}
        \lstick{$\ket{0}$} & \gate{X} & \gate{H} & \gate{H} & \meter{}\\
        \lstick{$\ket{0}$} & \qw & \qw & \qw & \meter{}
    \end{quantikz}};

    \node (left_4) at (0,4*\vspace) [anchor=north]{
    \begin{quantikz}
        \lstick{$\ket{0}$} & \gate{X} & \gate{H} & \gate{H} & \meter{}\\
        \lstick{$\ket{0}$} & \gate{X} & \qw & \qw & \meter{}
    \end{quantikz}};
    
    \node (right_1) at (1*\hspace,1*\vspace) [anchor=north]{
    \begin{quantikz}
        \lstick{$\ket{0}$} & \ctrl{1} & \ctrl{1} & \meter{}\\
        \lstick{$\ket{0}$} & \targ{} & \targ{} & \meter{}
    \end{quantikz}};

    \node at (1*\hspace,2*\vspace) [anchor=north]{
    \begin{quantikz}
        \lstick{$\ket{0}$} & \qw & \ctrl{1} & \ctrl{1} & \meter{}\\
        \lstick{$\ket{0}$} & \gate{X} & \targ{} & \targ{} & \meter{}
    \end{quantikz}};

    \node at (1*\hspace,3*\vspace) [anchor=north]{
    \begin{quantikz}
        \lstick{$\ket{0}$} & \gate{X} & \ctrl{1} & \ctrl{1} & \meter{}\\
        \lstick{$\ket{0}$} & \qw & \targ{} & \targ{} & \meter{}
    \end{quantikz}};

    \node (right_4) at (1*\hspace,4*\vspace) [anchor=north]{
    \begin{quantikz}
        \lstick{$\ket{0}$} & \gate{X} & \ctrl{1} & \ctrl{1} & \meter{}\\
        \lstick{$\ket{0}$} & \gate{X} & \targ{} & \targ{} & \meter{}
    \end{quantikz}};

    \begin{pgfonlayer}{background}
        \node[draw, fill=gnu_plot_1!30, dashed, inner sep=0pt, fit=(left_1) (left_4), label={[shift={(-1.0,0)}]left part}] (box_left) {};
        \draw[dashed] ($(box_left.north west)!0.25!(box_left.south west)$) -- ($(box_left.north east)!0.25!(box_left.south east)$);
        \draw[dashed] (box_left.west) -- (box_left.east);
        \draw[dashed] ($(box_left.north west)!0.75!(box_left.south west)$) -- ($(box_left.north east)!0.75!(box_left.south east)$);
        \node[draw, fill=gnu_plot_2!30, dashed, inner sep=0pt, fit=(right_1) (right_4), label={[shift={(1.0,0)}]right part}] (box_right) {};
        \draw[dashed] ($(box_right.north west)!0.25!(box_right.south west)$) -- ($(box_right.north east)!0.25!(box_right.south east)$);
        \draw[dashed] (box_right.west) -- (box_right.east);
        \draw[dashed] ($(box_right.north west)!0.75!(box_right.south west)$) -- ($(box_right.north east)!0.75!(box_right.south east)$);
    \end{pgfonlayer}

    \node (m1) at(0, 5.5*\vspace) {$M_{A1}$};
    \node (m2) at(\hspace, 5.5*\vspace) {$M_{A2}$};
    \node (m) at(0.5*\hspace, 5.5*\vspace) {$M_A=\dfrac{M_{A1}+M_{A2}}{2}$};

    \draw [arrow] (org) to [out=180, in=90] (left_1);
    \draw [arrow] (org) to [out=0, in=90] (right_1);
    \draw [arrow] (org) to [out=0, in=90] (right_1);
    \draw [arrow] (left_4) to [out=270, in=90] (m1);
    \draw [arrow] (right_4) to [out=270, in=90] (m2);
    \draw [arrow] (m1) to [out=0, in=180] (m);
    \draw [arrow] (m2) to [out=180, in=0] (m);
    
\end{tikzpicture}
    \caption{Outline of the approximated complete assignment matrix creation for the general mitigation protocol. We split the circuit into left (red) and right (blue) parts. Every part is extended with its inversion and run with all possible basis states to obtain the matrices $M_{A1}$ and $M_{A2}$ which we average to get $M_A$.}
    \label{fig_circuits_split}
\end{figure}

Reference~\cite{error_mit_jattana} proposes a general protocol to also consider the gate error. The method is used in various works to improve the quality of the results~\cite{citing_gem_1, citing_gem_2, citing_gem_3, citing_gem_4, citing_gem_5, citing_gem_6}. They split every circuit into two parts and append an inverted circuit to every part. To create the approximated complete assignment matrix, they prepare every basis state for both circuits, run the circuits, and perform a measurement. This yields two matrices $M_{A1}$ and $M_{A2}$ which are averaged to obtain the approximated complete assignment matrix $M_A=(M_{A1}+M_{A2})/2$. Figure \ref{fig_circuits_split} illustrates this procedure.

In order to have comparable errors in the calibration as in the actual circuit, it is essential to use the same gates. However, apart from the state preparation, an identity operation must be performed to create the complete assignment matrix. Simply inverting a circuit would result in a circuit with twice the depths and potentially significantly different error behavior. Hence, circuit splitting is a good compromise between running the original circuit and preserving the circuit depth~\cite{error_mit_jattana}.

The general method mitigates the error better than the previous one for most cases. However, Criterion~4 of the criteria in Section~\ref{sec_matrix_based} (practically realizable) is only met for circuits with few qubits, since we require $2^{n+1}$ calibration circuits to construct the complete assignment matrix. In Section~\ref{sec_sparse_matrix}, we propose a scalable mitigation method that is practically realizable for large qubit numbers.

\subsection{Assessment criteria for mitigation}
\label{sec_criteria_mitigation}
Reference~\cite{error_mit_jattana} introduces different scores to measure the success of a mitigation. $\Delta X$ is the distance between the mitigated and simulated frequencies, $\Delta V$ is the distance between measured and simulated frequencies. They are calculated as follows:
\begin{align}
    \Delta X &= \sqrt{\sum_{i=1}^{2^n}(x_i - s_i)^2}\\
    \Delta V &= \sqrt{\sum_{i=1}^{2^n}(v_i - s_i)^2}\\
\end{align}
where $x$ are the mitigated, $v$ the measured and $s$ the simulated frequencies. They also introduce the score $\Delta Q$ which is the difference between $\Delta X$ and $\Delta V$:
\begin{equation}
	\Delta Q = \Delta V - \Delta X.
\end{equation}

If the mitigated result is closer to the simulated than the measured one, $\Delta Q$ is positive, otherwise negative. Higher values of $\Delta Q$ indicate that the mitigation worked better. In this work, we use $\Delta Q$ to measure the success of mitigation and assess Criterion~1.
\section{Scalable error mitigation}
\label{sec_sparse_matrix}
\begin{figure}[htbp]
	\centering
	\import{figures/heatmap_average_simulation}{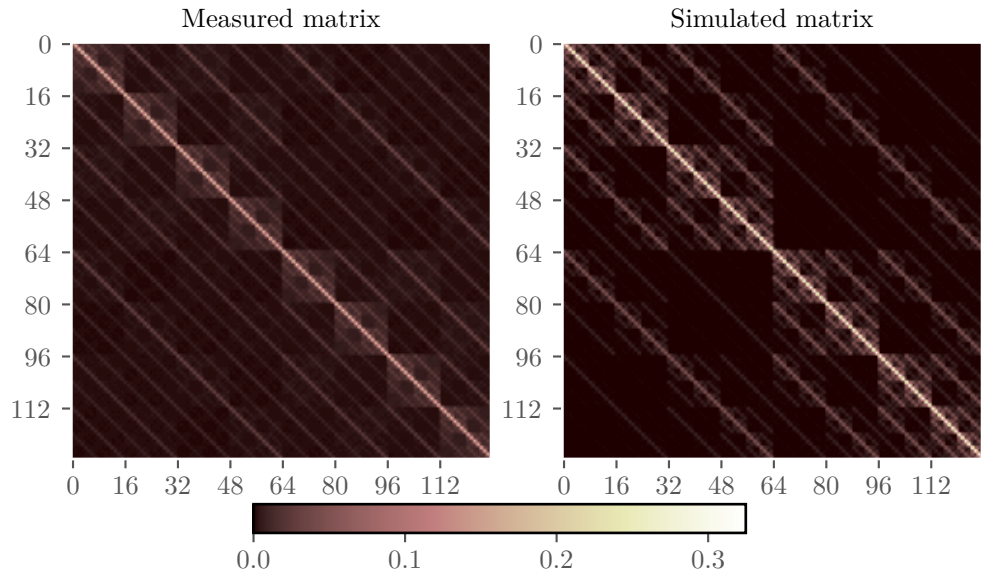}
	\caption{Left: average of 103 complete assignment matrices of randomly created circuits with 7 qubits plotted as a heat map. The matrices are obtained with the GEM method. The values of the measured frequencies are color-coded in the columns. Lighter colors represent larger values. Right: complete assignment matrix for 7 qubits obtained from a Monte Carlo simulation. We prepare every possible state of the $z$ basis and flip every qubit with $p=0.15$. The average of $10^5$ such runs is plotted as a heat map. We can see similar patterns in both matrices.}
	\label{fig_heatmap_average_simulation}
\end{figure}

In this section, we introduce a mitigation method that improves the general error mitigation method to make it suitable for circuits with a large number of qubits. Reference~\cite{error_mit_spam_3} shows that the measurement errors are distributed over states that are close in Hamming distance to the actual states, since we can assume a bitflip error model as a first approximation for the state preparation and measurement. This leads to many values in the complete assignment matrix being negligible for mitigation. We observe the same in our experiments. Figure~\ref{fig_heatmap_average_simulation} shows on the left side the average of the assignment matrices for 103 experiments with 7 qubits and randomly created circuits. The matrices are obtained with the GEM method. On the right side we see a matrix from a Monte Carlo simulation where every bit is flipped with probability $p=0.15$. We see similar patterns of diagonal lines and squares with larger values, while most values are close to zero, for both matrices. We conclude that the errors in the whole circuit, not just in the state preparation and measurement, behave approximately like the bitflip error model, and that we can neglect values with a large Hamming distance for mitigation. We can therefore expect to omit most values in the matrix without significant change in $\Delta Q$.

A problem to this approach is that we cannot simply define a threshold for the matrix and set every value below it to zero, since we first face the problem that we cannot construct the full matrix without an exponential number of calibration circuits. The solution is to construct a smaller matrix directly. Our strategy is to select $k$ relevant states in $V$ and only mitigate those, rather than mitigating all. For a reasonable quantum algorithm, we can assume that $k$ grows no more than polynomially with the number of qubits. If it grew exponentially, we would require an exponential number of shots, which would violate Criterion~4. The procedure is as follows, considering we have run a circuit of depth $D$:

\begin{enumerate}[(1)]
    \item Run the original circuit to get the output frequencies.
    \item Choose the $k$ states with the largest frequencies in the obtained results and store them in the set $P=\{p_0, p_1, \ldots, p_{k-1}\}$, where $p_i$ is the decimal representation of the state. The order of $P$ does not matter.
    \item If $D$ is even, split the circuit after $D/2$ gates, if odd, after $(D-1)/2$ gates. Invert the respective sub-circuits and append them to the original halves to get two calibration circuits.
    \item Construct $k$ calibration circuits that prepare the states in $P$.
    \item Combine the circuits from the previous steps to get $2k$ calibration circuits and run these calibration circuits on the device.
    \item Use the measured frequencies to create the matrices $M'_{A1}$ and $M'_{A2}$ and combine them into a matrix $M'_A=(M'_{A1}+M'_{A2})/2$. The matrices $M'_{A1}$ and $M'_{A2}$ are constructed similar as shown in Section~\ref{sec_general_mitigation}. The difference is that $M'_{A1}$ and $M'_{A2}$ are $k \times k$ matrices containing only elements $m_{p_i p_j}$, where $p_i, p_j \in P$.
	\item Perform the error mitigation by minimizing $f(x)$ with
	\begin{equation}
		f(x) = \sum_{i=0}^{k-1}(v'_i - (M'_A X')_i)^2,
	\end{equation}
	and the constraints $0\leq x_i \leq 1$ and $\sum x_i = 1$ for all $i$. $V'$ is a vector containing the measured frequencies $v_{p_0}, v_{p_1}, \ldots, v_{p_{k-1}}$ of the original circuit, and $X'$ contains the mitigated frequencies $x_{p_0}, x_{p_1}, \ldots, x_{p_{k-1}}$. Adhering to the constraints means that the mitigated frequencies $X'$ sum up to one and the frequencies of all other states are zero.
\end{enumerate}    

The following example illustrates the procedure: Assuming that we have a 3 qubit circuit with depth $D$ and $k=3$. We first run the circuit to get $V$ (step~1). We then chose the three states with the largest frequencies from $V$ and store their decimal representation in $P$ (step~2). For this example we assume these states to be $\ket{001}$, $\ket{011}$ and $\ket{111}$, therefore $P=\{1,3,7\}$. We split the circuit, append the inverses (step~3) and run both halves for the input states $\ket{001}$, $\ket{011}$ and $\ket{111}$ (step~4 and 5). We create the matrix $M'_A$ as
\begin{equation}
	M'_A=\begin{pmatrix}
		m_{11} & m_{13} & m_{17}\\
		m_{31} & m_{33} & m_{37}\\
		m_{71} & m_{73} & m_{77}
	\end{pmatrix}
\end{equation}
(step~6) and perform mitigation (step~7).

Note that the columns in $M'_A$ do not necessarily add up to 1, like they do in $M_A$. As $k\to2^n$, $\sum_{i=0}^{k-1} M'_{Aij} \to 1$ for all $j = 0, 1, \dots, k-1$. Another approach would be to renormalize the columns of $M'_A$ and $V'$ before performing the mitigation. Further work could explore whether this renormalization would be beneficial for mitigation. We do not explicitly include states that are close in Hamming distance to the actual states, but we implicitly include them if their frequencies are large, since they then belong to the $k$ states with the largest frequencies.

\subsection{State selection}
\label{sec_state_selection}
In the example of the last section, we chose the number of states $k$ arbitrarily. Here we discuss some considerations. If we choose $k$ too large, the overhead for error mitigation becomes impractical, violating Criterion~4. If $k$ is too small, the mitigation does not improve the results, violating Criterion~1. We now propose a strategy for choosing $k$ that works independent of the circuit depths and does not rely on knowledge about the simulated result of the circuit (see criteria 2 and 7 in Section~\ref{sec_introduction}).

\begin{figure}[htbp]
    \centering
    \input{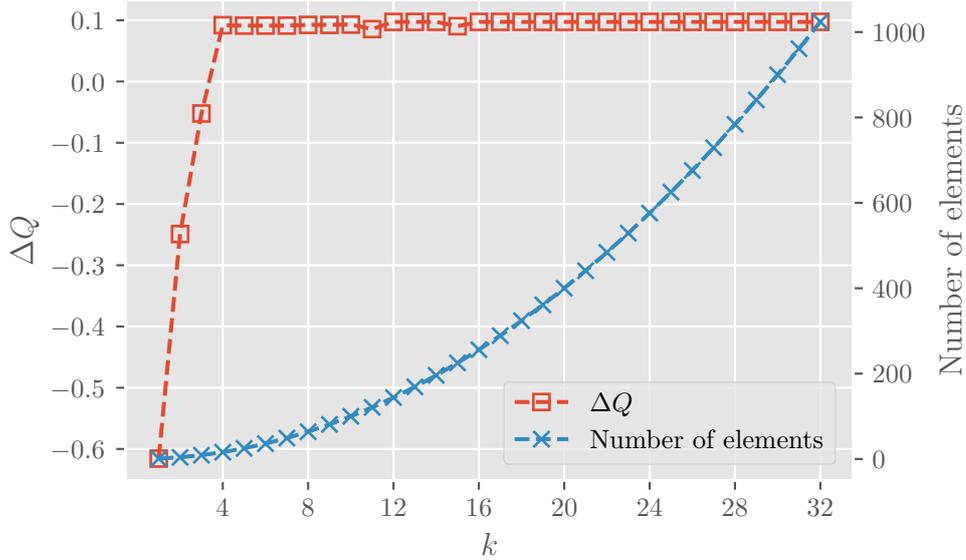}
    \caption{$\Delta Q$ and matrix size plotted for different $k$. For $k<4$, $\Delta Q$ increases strongly for increasing $k$. For $k>4$ there is not much change in $\Delta Q$ for increasing $k$. Therefore, it is optimal to choose $k=4$ for this example.}
    \label{fig_delta_q_vs_k}
\end{figure}

Instead of choosing any fixed $k$, we propose an iterative approach, where $k$ is increased until considering more states does not improve the result anymore. In our experiments, this is often the case if $k$ equals the number of non-zero states in the simulated distribution. One example is depicted in Figure~\ref{fig_delta_q_vs_k}. It shows $\Delta Q$ plotted for increasing $k$ of a circuit with 5 qubits and 4 non-zero states in the simulated result. We see the steep increase in $\Delta Q$, plotted in red, from a negative to a positive value until $k=4$. The number of elements in the required matrix is additionally plotted in blue. We see that $k=4$ is optimal between small matrix size and large $\Delta Q$ for this example. Increasing $k$ further does not increase $\Delta Q$. This is the typical behavior observed in our experiments.

In Section~\ref{sec_criteria_mitigation}, we introduce $\Delta Q$ as the difference between $\Delta V$ and $\Delta X$, which are defined as the distance of the measured and the mitigated frequencies to the simulated ones. Since we used the simulated frequencies, we can use $\Delta Q$ only for the validation. We cannot use $\Delta Q$ to find the most optimal $k$. To overcome this problem, we introduce a different score $\Delta R$, which is the distance between the measured and mitigated frequencies:
\begin{equation}
    \Delta R = \sqrt{\sum_{i=1}^{2^n}(v_i - x_i)}.
\end{equation}
This new score does not need results from a simulator in accordance with criterion 7 of the ideal criteria mentioned in Section~\ref{sec_introduction}.

\begin{figure}[hbp]
    \centering
    \input{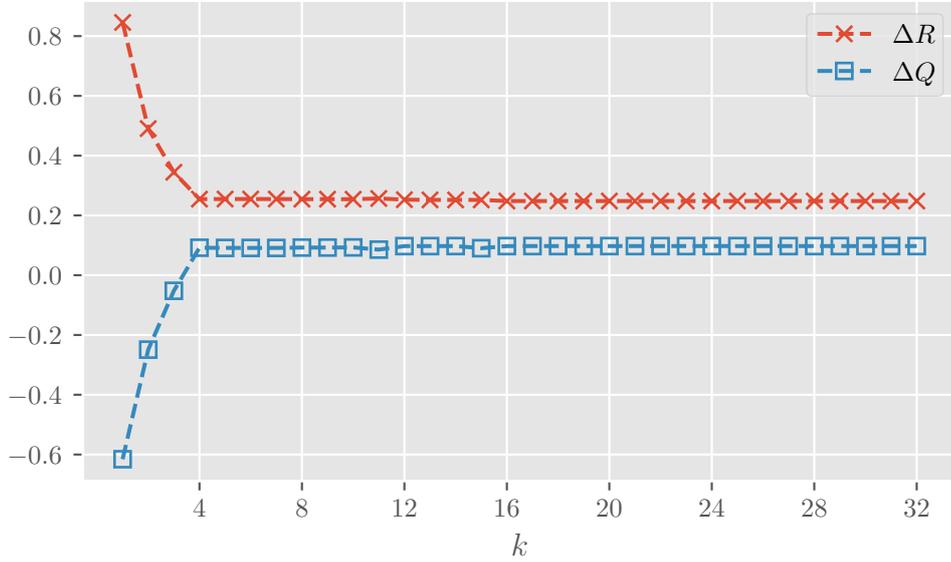}
    \caption{$\Delta R$ and $\Delta Q$ of a 5 qubit experiment plotted as function of $k$. We see a change in $\Delta R$ if $\Delta Q$ changes.}
    \label{fig_delta_r_vs_k}
\end{figure}

Since the distance between the measured and simulated frequencies is constant for different $k$, we know that $\Delta R$ changes if $\Delta Q$ changes. We simply relate measured and mitigated frequencies directly to each other instead of relating them to the simulated ones. Figure \ref{fig_delta_r_vs_k} shows that this assumption holds for our experiments: If $\Delta Q$ changes, $\Delta R$ changes as well.

To perform the mitigation, we set a threshold $t$ and a maximum number of states $k_{\text{max}}$. We iteratively consider more and more states and calculate $\Delta R$, until either the change in $\Delta R$ is below $t$, or $k=k_{max}$. The algorithm works as follows:

\begin{algorithmic}[1]
\State Run original circuit
\State $k \gets 1$
\State $\Delta R \gets 0$
\State $\Delta R_{\text{prev}} \gets t$
\While{$k \le k_{\text{max}} \operatorname{\&\&} |\Delta R_{\text{prev}} - \Delta R| \leq t$}
    \State Choose state with the $k^{\text{th}}$ largest frequency and run calibration circuits
    \State Prepare $M'$ with $k\times k$ elements
    \State Perform error mitigation with $M'$
    \State Calculate new $\Delta R$
    \State $\Delta R_{\text{prev}} \gets \Delta R$
    \State $k \gets k + 1$
\EndWhile
\end{algorithmic}

\section{Experimental results}
\label{sec_experimental_results}

We use randomly generated circuits to test how well the error mitigation works compared to the general method~\cite{error_mit_jattana}, and how large the matrices are for each procedure. The generation process for the randomly generated circuits must meet the following criteria:

\begin{enumerate}[a.]
	\item An arbitrary width of the circuit.
	\item The depths of the circuit after transpilation to the device-specific gate set must be settable.
	\item The probability distribution when sampling from the circuit must contain a preferable number of states. It must be possible to set whether more or fewer of these states are present.
\end{enumerate}

To meet criteria~a and b, we always generate circuits for one specific device. From the device description we obtain the number of qubits which determines the maximum possible width and the set of native gates, from which we pick randomly until the desired depths are reached. By using the native gate set of the device, we ensure that no gate is decomposed into multiple native gates during compilation. We also disable any compiler optimization that could lead to the cancellation of certain gates. 2-qubit gates are only applied on physically neighboring qubits of the actual device to avoid swap gates. For gates with an arbitrary rotation angle like the $\text{R}_\text{z}$ gate, we set the angle randomly to a value in the range $[0, 2\pi)$. To meet Criterion~c, we set the number of SX gates in the circuit, since they are required to create entanglement along with CX gates and therefore influence the number of nonzero states.

\begin{figure}[ht]
	\centering
	\input{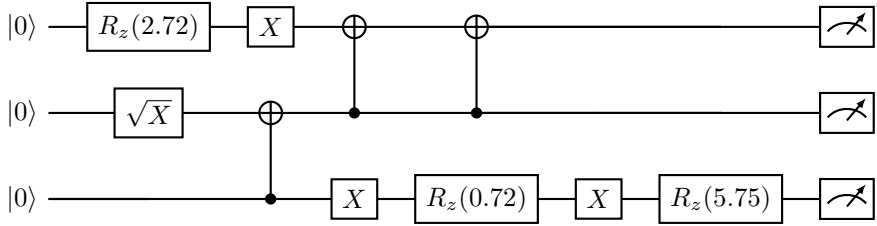}
	\caption{Randomly created circuit with width 3 and depth 6 for the IBM device Jakarta. The circuit can be executed directly on the device without further transpilation.}
	\label{fig_circuit_example}
\end{figure}

Figure \ref{fig_circuit_example} shows an example of a circuit created for IBMQ Jakarta device with width 3 and depth 6. There are two successive CX gates applied on the top most qubits. Since the CX gate is self-inverse, they could be canceled out by an optimizer. Therefore, it is important to run the circuit unoptimized to keep the circuit depths as intended.

We carry out all experiments on IBMQ superconducting systems \cite{ibm_devices, ibm_devices_old}, more precisely on the systems \emph{IBMQ Jakarta}, \emph{IBMQ Lagos}, \emph{IBMQ Nairobi} and \emph{IBMQ Kyiv}. For programming the devices we use IBMs Qiskit framework~\cite{qiskit}.

In our experiments, we test two different aspects: first we test how the truncated matrix approach compares to the general mitigation protocol~\cite{error_mit_jattana}. To achieve this goal, we construct circuits with a small number of qubits, for which it is possible to build the whole matrix. Second, we test if the method works well with a large number of qubits for which we cannot construct the full matrix. We only test a very simple circuit to prove that the methods works on a large scale.

\subsection{Tests on simulatable devices}
\begin{figure}[htbp]
    \centering
    \input{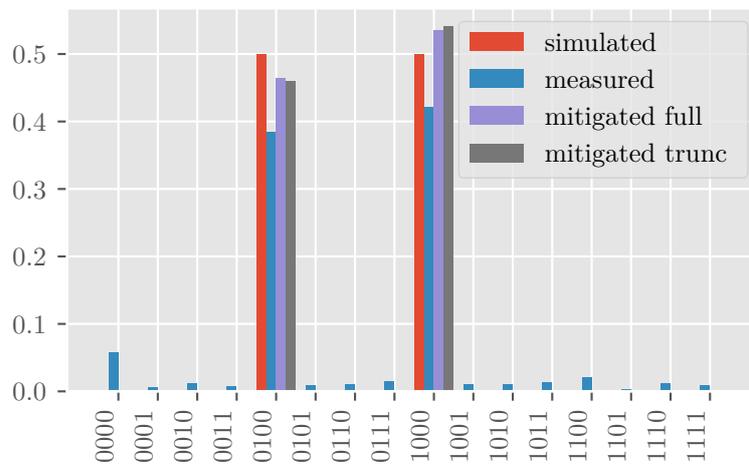}
    \caption{Example of one experiment with a 4 qubit circuit. The plot shows the simulated and measured results as well as the mitigated results with full and truncated matrix.}
    \label{fig_mitigation_single_circuit}
\end{figure}

To show that the mitigation works, we test numerous randomly created circuits. These circuits have widths between 2 and 7 qubits and depths between 10 and 140 gates, making it possible to simulate them completely to obtain the simulated results. We compare the measured and mitigated results against the simulation. For the mitigation, we use the full matrix and truncated matrix approach to compare both methods. Figure~\ref{fig_mitigation_single_circuit} shows one example of an experiment with 4 qubits. The red bars show the result of the simulation. We see that the measured results (blue bars) differ a lot from the simulation. The mitigated results with full matrix (purple) as well as with truncated matrix (gray) are both much closer to the simulated result. For this example the mitigation with the full matrix ($\Delta Q = 0.106$) works slightly better than with the truncated matrix ($\Delta Q = 0.098$).

\begin{figure}[htbp]
    \centering
    \input{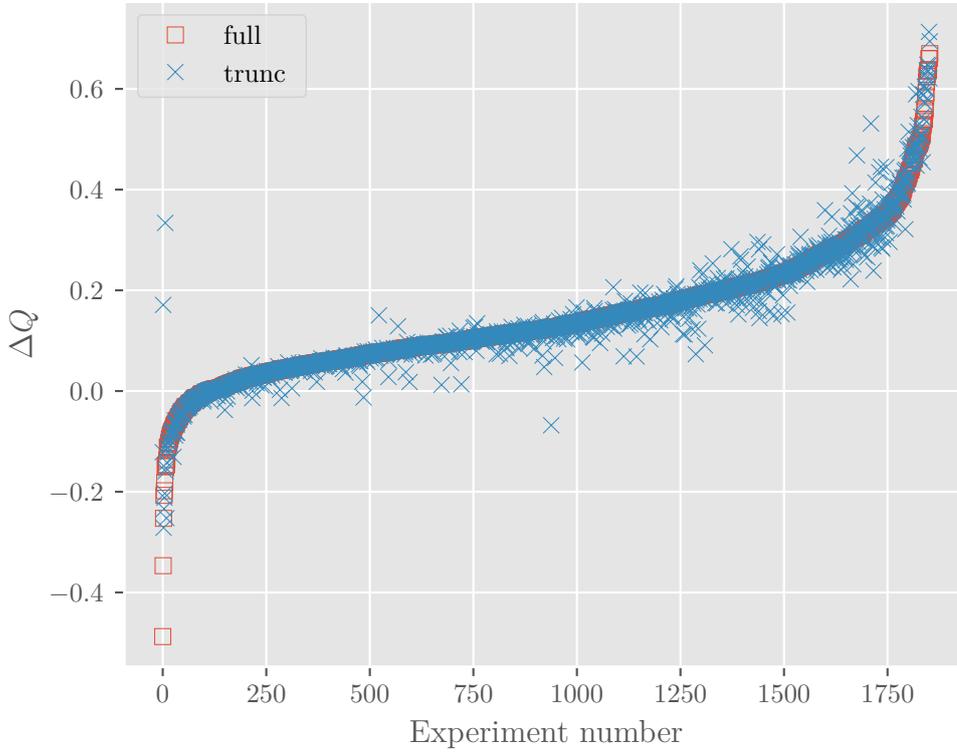}
    \caption{$\Delta Q$ of the mitigation with full and truncated matrices for 1853 experiments ranging between 2-7 qubits and 10-140 gates depths. The results are sorted by the $\Delta Q$ of the full matrix in the ascending order.}
    \label{fig_delta_q_diff_all}
\end{figure}

\begin{figure}[htbp]
    \centering
    \input{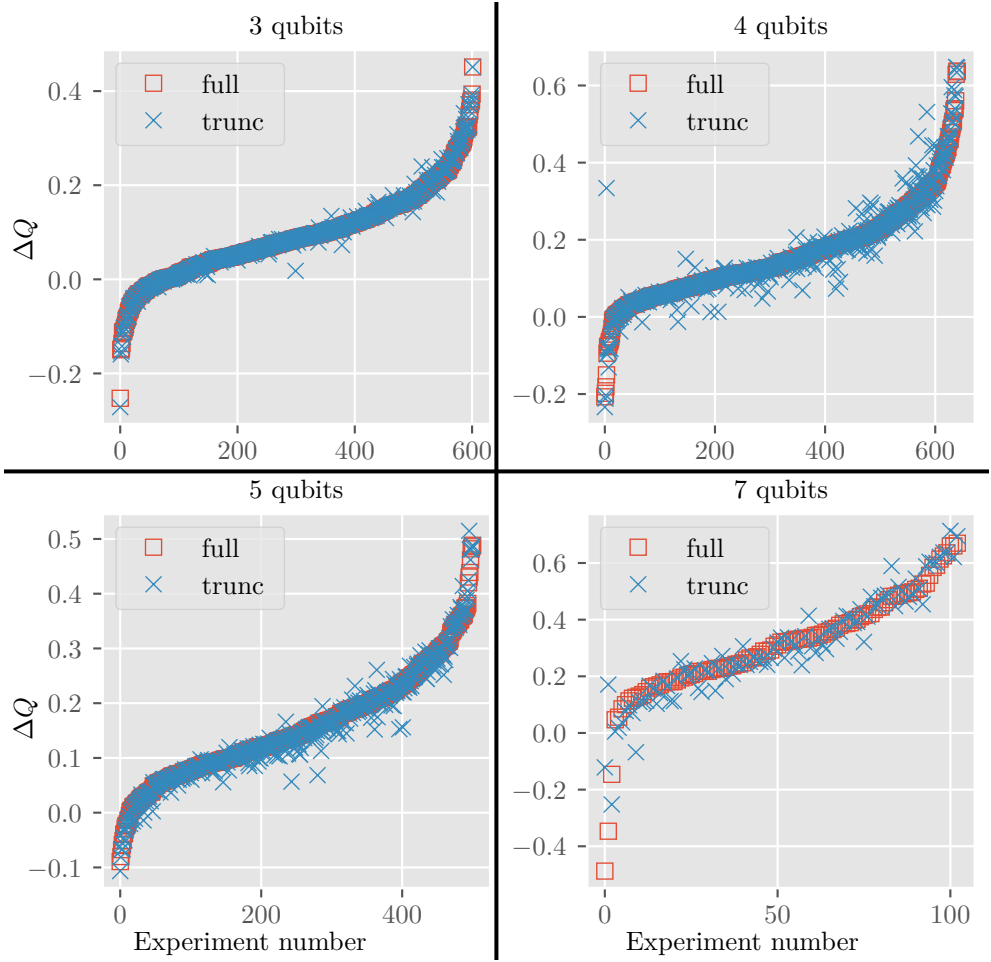}
    \caption{$\Delta Q$ of the mitigation with full and truncated matrix for 3, 4, 5, and 7 qubits. The results are sorted by the $\Delta Q$ of the full matrix in ascending order.}
    \label{fig_delta_q_diff_single}
\end{figure}

Since our truncated matrix approach is a heuristic method and the results a quantum computer yields as well as the errors are stochastic, we cannot draw conclusions from a single result, but need to look at multiple experiments. Therefore, we run lots of randomly created circuits, calculate $\Delta Q$ and average for all experiments. In total, we run 1853 such experiments. Figure~\ref{fig_delta_q_diff_all} shows the result of these tests with the experiments sorted by ascending $\Delta Q$ of the full matrix. We see that the $\Delta Q$ of the truncated matrix mitigation (blue crosses) follows the trend of the full matrix (red squares) with some upward and downward outliers. In Figure~\ref{fig_delta_q_diff_single} we see the same plots for 3, 4, 5 and 7 qubits separately. There are some outliers on the left and right side where the mitigation worked very poorly or very well. Since this is a heuristic method, these can occur randomly. However, we often get a very large $\Delta Q$ when there is only one non-zero state in the output distribution and a negative $\Delta Q$, with many non-zero states. Note that the experiments for 2 qubits are not shown here. Table~\ref{tab_full_vs_sparse_results} lists the average $\Delta Q$ of truncated and full matrix mitigation for different number of qubits. We see, that there is only a slight difference.

\begin{table}[htbp]
    \centering
    \caption{Average $\Delta Q$ of mitigation with full and truncated matrix for different number of qubits. We see that the truncated matrix works well for circuits of different widths.}
    \label{tab_full_vs_sparse_results}
    \begin{tabular}{|l|r|r|r|r|r|r|}
        \hline
        Number of qubits         & 2 & 3 & 4 & 5 & 7 & total\\
        \hline
        Number of experiments    & 6 & 602 & 641 & 501 & 103 & 1853 \\
        \hline
        Average $\Delta Q$ full matrix   & 0.059 & 0.094 & 0.160 & 0.157 & 0.309 & 0.146\\
        \hline
        Average $\Delta Q$ truncated matrix & 0.059 & 0.095 & 0.160 & 0.154 & 0.308 & 0.145\\
        \hline
    \end{tabular}
\end{table}


\subsection{Storage requirements}
\begin{figure}[htbp]
    \centering
    \input{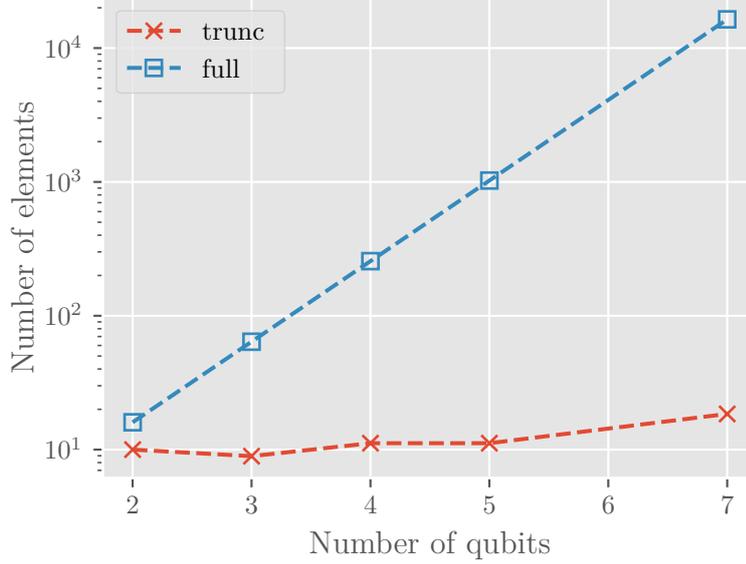}
    \caption{Average storage requirement of the full and truncated matrix for different numbers of qubits.}
    \label{fig_matrix_size}
\end{figure}

In the previous section, we showed that the mitigation with a truncated matrix works comparably well as with the full matrix. This is however only meaningful if the matrix is much smaller than the full matrix. In Figure~\ref{fig_matrix_size}, we plot the number of elements of the truncated and full matrix. We see that the full matrix size grows exponentially with the number of qubits, as expected. The average size of the truncated matrix is independent of the number of qubits. It is dependent only on the number of states taken into consideration for mitigation.

\subsection{Large scale test}
\begin{figure}[htbp]
	\centering
	\input{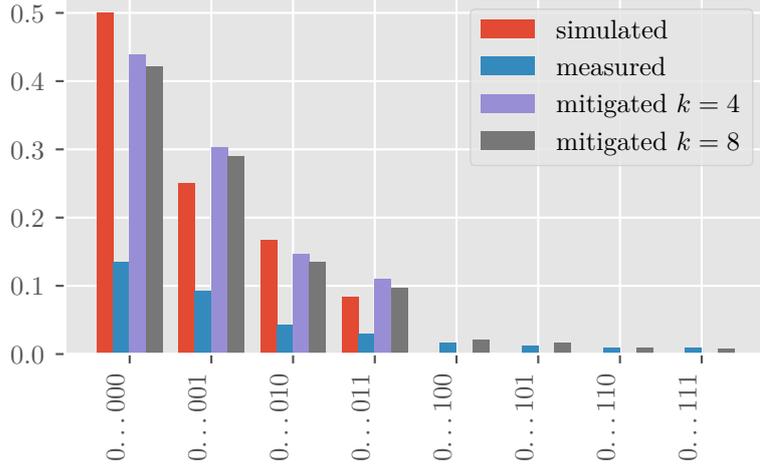}
	\caption{Large scale experiment with 100 qubits and two $R_x$ gates on the first two qubits. The mitigation improves the result for $k=4$ as well as $k=8$.}
	\label{fig_mitigation_single_circuit_large}
\end{figure}

\begin{figure}[htbp]
	\centering
	\input{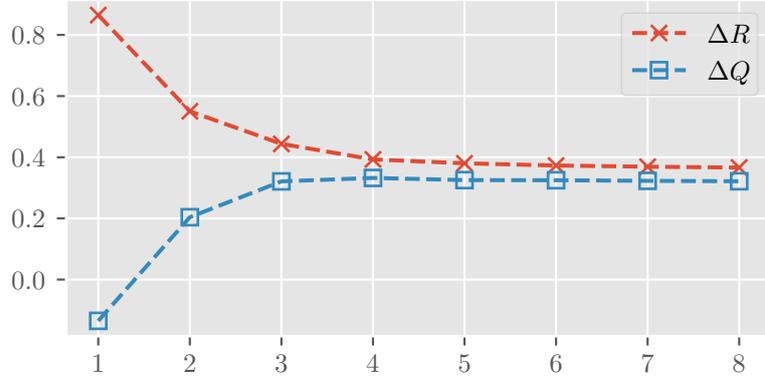}
	\caption{$\Delta R$ and $\Delta Q$ of a 100 qubit experiment plotted as function of $k$. We see a change in $\Delta R$ if $\Delta Q$ changes.}
	\label{fig_delta_r_vs_k_large}
\end{figure}

The purpose of the large scale test is to show that our method works on circuits of sizes that matrix mitigation with the full matrix cannot handle. We use 100 qubits for our test. On the first two qubits we perform rotations around the $x$ axis to get 4 output states with different weights on the qubits 0\ldots00 - 0\ldots11. We run calibration circuits for the 8 states with the largest frequencies in the measured output. Figure~\ref{fig_mitigation_single_circuit_large} shows the output distribution with simulated, measured and mitigated results. The mitigated results are shown for $k=4$ and $k=8$. In Figure~\ref{fig_delta_r_vs_k_large} we see  $\Delta Q$ and $\Delta R$ plotted as a function of $k$. They show the same behavior as for small circuits: $\Delta Q$ increases until $k$ is the number of non-zero states in the simulated output (4 in this case) and doesn't change significantly afterwards. $\Delta R$ decreases sharply until $k=4$. This experiment shows that our method can be applied to an arbitrary number of qubits that would traditionally be out of reach of non-scalable methods.

\section{Conclusion}
\label{sec_conclusion}

We introduced a heuristic method for scalable matrix based error mitigation and tested it on real superconducting hardware from IBM. Our experiments demonstrated, that it performs comparably well to a matrix-based method using the full complete assignment matrix, but requires a fraction of the storage and calibration runs. We also demonstrated that the method works for large qubit numbers.

In future work, it may be interesting to systematically compare different scalable methods to determine which method works best under which conditions. It may also be useful to use several methods in sequence to achieve a better overall result than with individual methods. As soon as quantum processors work well enough to implement error correction algorithms in a meaningful way, it will be interesting to explore the combination of error correction and error mitigation. 

We saw for example in Figure~\ref{fig_mitigation_single_circuit_large}, that mitigation did not work perfectly. So far, we have only created the complete assignment matrix using calibrations in the $z$ basis. It will be interesting to investigate whether other choices of basis can improve the mitigation.

\clearpage
\begin{appendices}




\end{appendices}


\bibliography{sn-article}

\end{document}